# Rhythm and form in music: a complex systems approach

Blas Kolic[1,2], Mateo Tonatiuh Rodriguez-Cervantes[3,†], Pablo Padilla-Longoria[3], Francis Knights[4]

[1]Mathematical Institute, University of Oxford, Oxford CP, UK
[2]Institute for New Economic Thinking, University of Oxford, Oxford CP, UK
[3]Institute for Applied Mathematics and Systems (IIMAS), UNAM, Ciudad Universitaria, Mexico City 04510, México
[4]Fitzwilliam College, University of Cambridge, Cambridge CB3 0DG, UK

[†]mateotonatiuh@ciencias.unam.mx

> "Music is similar to language in that it is a temporal succession of articulated sounds that are more than just sound. They say something, often something humane. The higher the species of music, the more forcefully they say it. The succession of sounds is related to logic; there is a right and a wrong. But what is said cannot be abstracted from the music; it does not form a system of signs."
> —Theodor W. Adorno, *Music, Language and Composition* [5]

## ABSTRACT

There has been an everlasting discussion around the concept of form in music. This work is motivated by such debate by using a complex systems framework in which we study the form as an emergent property of rhythm. Such a framework corresponds with the traditional notion of musical form and allows us to generalize this concept to more general shapes and structures in music. We develop the three following metrics of the rhythmic complexity of a musical piece and its parts: 1) the *rhythmic heterogeneity*, based on the permutation entropy [2], where high values indicate a wide variety of rhythmic patterns; 2) the *syncopation*, based on the distribution of on-beat onsets, where high values indicate a high proportion of off-the-beat notes; and 3) the *component extractor*, based on the communities of a visibility graph of the rhythmic figures over time [3], where we identify structural components that constitute the piece at a (to be explained) perceptual level. With the same parameters, our metrics are comparable within a piece or between pieces.

**Keywords**: Rhythmic complexity, information theoretical methods, network-based methods, stylistic analysis

## INTRODUCTION

To set the preliminary musical notions, we will depart from Adorno's music comparison to language in [5]: "a temporal succession of articulated sounds that are more than just sound", but generalizing the concept of *sound* to *sound event*. Thinking in a general way, the events that constitute a musical piece are sounds (of different kinds and with various features, produced by one or several sources) or silences (of some or all the sources at the same or different time), arranged one after the other from the beginning till the end of a piece. The way these events are organized is called rhythm and plays a crucial role in the structure of all music.

We can say, in the most general and abstract way, that to talk about form in music is to talk about the relation of the sound events between each other (e.g., melodic cells, motives, phrases, themes, etc.) and between groups (and groups of groups) of events (e.g., song, rondo, sonatina, sonata, figuration, fugue, canon, etc.). We can compare two visions of this concept in different times, one from the Romantic era, by A. B. Marx [4], when this concept was emerging: "[...] Form in music is thus nothing other than the shaping and hence determination of content that is originally shapeless and undetermined but lies ready in the spirit, eagerly awaiting musical shape, and only then through shaping, through form becoming music. [...]". And another from the XX century composer and music theorist James Tenney [1]: "FORM. In the most general sense: shape (contour, the variation of something in space or time), and structure (the disposition of parts, relations of part to part, and of part to whole). In music, shape results from changes in some attribute or parameter of sound in time, while structure has to do with various relations between sounds and sound-configurations, at the same or at different moments in time. The wFirst, we must define what a note "on the beat" is and then clarify what we mean by *distance*. We say a note is *on-the-beat i*f it stats at an integer multiple of the *beat length, b*, with respect to the beginning of a measure. Typically, the time signature indicates the beat length. For example, the beat length is typically the quarter note for pieces in 4/4.ord is often used in the more restricted sense of a fixed or standard scheme of relationships (e.g., "sonata-form"). Still, this definition of form is of little use in a study of music in the 20th century, which has tended to break away from such fixed patterns, yielding a fantastic variety of new forms. To deal with this variety, our basic definition of form must be as broad as possible, and several new terms will have to be developed.".

While the traditional way to understand form is well established for western music from the renaissance to the late-romantic (in most cases), the discussion about form is still open and asks for innovative ways of interpretation in the case of non-western and post-1900's western music. With this in mind, we describe a mathematical but intuitive framework to quantify the rhythmic complexity of a musical piece and its parts. Such a framework lets us see form as an emergent property of rhythm with metrics that correspond with the traditional notion of musical form but also allows us to generalize this notion to other shapes and structures.

## METHODOLOGY

The complexity of a form is necessarily tied to how we perceive it. Some authors argue that stimuli (of rhythms in music) that are too predictable are boring, while those unpredictable are unrewarding [8, 7]. Thus, a measure of rhythmic heterogeneity would be a natural candidate for its complexity. However, studies show that heterogeneity alone cannot explain how people perceive and feel rhythm; it also depends on syncopation, groove, and structure [9]. Such structure is linked with a perception phenomenon called subjective grouping [7], in which the lengthening of sounds plays a decisive role. In this paper, we introduce quantitative metrics of rhythmic complexity that are intuitive from a perceptual and a mathematical perspective. In what follows, we present the permutation entropy, disequilibrium, and modular visibility of a *rhythm* as a proxy of its rhythmic heterogeneity, syncopation, and structure, respectively.

We start by defining how we measure rhythm quantitatively: the rhythm of a voice or aggregate of voicings of a piece consists of its *inter-onset intervals (IOI)* vector. The IOI between two successive notes is the time between each hit, so for instance, if we have that ♩ = 30 and the following rhythm: (♩ ,♩ ,♪,♪,♩ ), then IOI = (2,2,1,1,2).

To show how our methods work, we will consider the first four measures from Mozart's Piano Sonata no.11 in A Major, K331. This piece is interesting because although it is short, it has all the typical elements of the sonata form.

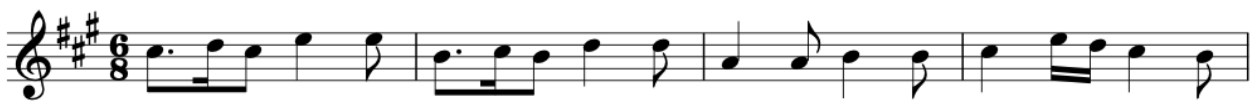

Figure 1. Four first measures from Mozart's Piano Sonata no.11 in A Major, K331.

**Rhythmic Heterogeneity**

We quantify the *rhythmic heterogeneity* of a piece with the *permutation entropy* of its IOI vector. The entropy indicates the level of uncertainty of the possible outcomes of an event; it is minimum when we know what the outcome of the event will be (imagine you are halfway through an AC/DC song where the bass has been playing only quarter notes; you can almost be sure that it will keep playing quarter notes for the rest of the song), and it is maximum when we cannot discern between all of the possible outcomes (now you are listening to the trumpet solo of a free-jazz improvisation; you will hardly know what rhythmic patterns to expect next).

The permutation entropy quantifies the uncertainty of sequences of notes rather than the uncertainty of the duration of the notes alone. Imagine you take a sequence of two notes. The first note can either 1) have the same duration as the second, 2) be shorter than the second, or 3) be longer than the second; these are all the possible permutations of two-note sequences. By taking all the two-note sequences in a piece, we can count how many of them fall into each of the three categories described above; i.e., we can build its distribution. If all of these sequences fall into the same category, say, the first note is longer than the second, then the permutation entropy is minimal. However, if the sequences were evenly split between the three categories, the permutation entropy, and therefore the uncertainty about the sequences, would be maximal. We can construct the above process for note sequences of any length $D$, so $D$ is a parameter of the permutation entropy. Mathematically, we define the rhythmic heterogeneity following [2] as

$$\mathcal{H} = -\frac{1}{\log N_D} \sum_\pi p_\pi \log p_\pi \qquad \text{(Equation 1)}$$

where $N_D$ is the number of possible sequences of size $D$, and $p_\pi$ is the frequency of sequences in category $\pi$. By dividing $\mathcal{H}$ by $\log N_D$, we ensure that the permutation entropy is bounded between 0 and 1. See Figure 2 for a graphical representation of the heterogeneity metric.

**Syncopation**

We obtain a proxy of the *syncopation* of a piece by measuring the distance between the distribution of *off-the-beat* notes and the distribution in which all the notes would have landed *in the beat*. First, we must define what a note "on the beat" is and then clarify what we mean by *distance*. We say a note is *on-the-beat* if it stats at an integer multiple of the *beat length*, $b$, with respect to the beginning of a measure. Typically, the time signature indicates the beat length. For example, the beat length is typically the quarter note for pieces in 4/4. We say that a note is *off-the-beat* if it is not on-the-beat. For every measure in the piece, we take the proportion of off-the-beat notes, giving us a set of all the off-the-beat proportions in all the measures in the piece. The next step is constructing the off-the-beat distribution, $P_{\text{off}}$, by first binning the interval $[0, 1]$ in equally spaced bins and then assigning each proportion to its corresponding bin. If $P_{\text{off}}$ has most of its weight near 1, it means that most notes are off the beat. We define *syncopation* as the cost of transforming the off-the-beat $P_{\text{off}}$ distribution into a perfect on-the-beat distribution $P_*$ in which every note of every measure lands on the beat. We quantify such a cost with the *earth mover's distance* [10], or EMD. Informally, suppose the distributions are interpreted as two different ways of piling up a certain amount of dirt. In that case, the EMD is the minimum cost of turning one pile into the other. Here, the cost means the amount of dirt moved times the distance by which it is moved. Thus, the EMD meets our notion of syncopation: it is 0 when every note in the score is on the beat, and it is maximal where every note is off the beat; every other configuration falls in between. Mathematically,

$$\mathcal{Q} = \frac{1}{N_{\text{bins}} - 1} EMD(P_{\text{off}}, P_*) \qquad \text{(Equation 2)}$$

where $N_{\text{bins}}$ is the number of bins and $N_{\text{bins}} - 1$ is the maximum value the EMD can take, so that $\mathcal{Q} \in [0, 1]$. See Figure 2 for a graphical representation of the syncopation metric.

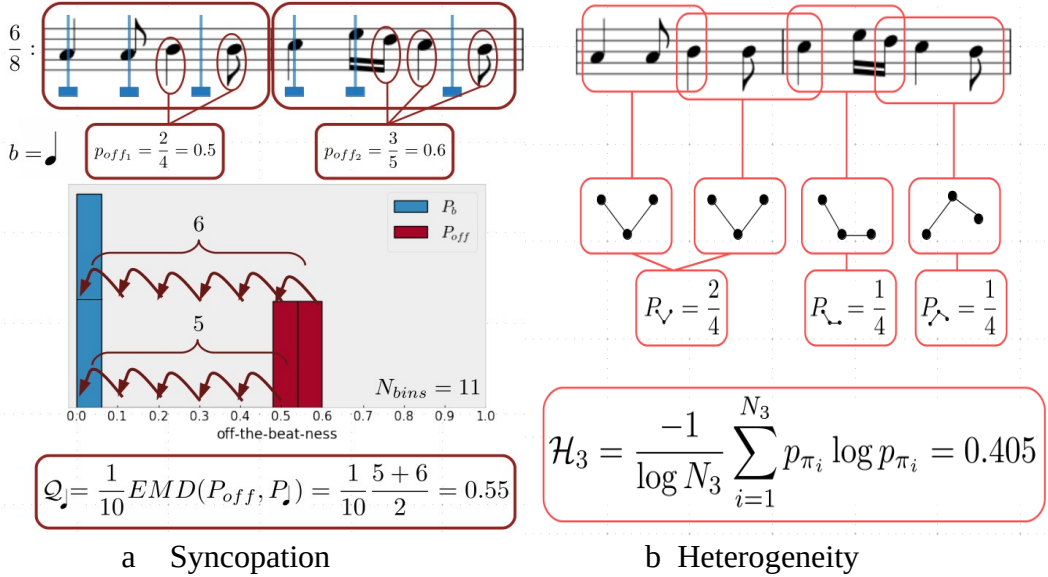

Figure 2. Schematic representation of the syncopation, $\mathcal{Q}$ (left), and the rhythmic heterogeneity, $\mathcal{H}$ (right), for measures 3 and 4 of Mozart's Sonata. *a)* The syncopation metric quantifies the distance between the distribution of *off-the-beat* notes and the distribution where all notes land *on-the-beat* for each measure. In the given example, 2 out of 4 notes (i.e., 0.5) on the first measure land off the beat, while 3 out 5 notes (i.e., 0.6) on the second measure land off the beat, resulting in the distribution $P_{\text{off}} = (p_{0.5} = 0.5, p_{0.6} = 0.5)$. The EMD from $P_{\text{off}}$ to $P_* = (p_0 = 1)$ is 11/2, resulting in $\mathcal{Q} = 0.55$. *b)* The heterogeneity metrics the variation of rhythmic phrases of length $D$ using the permutation entropy of those phrases. In the example, we take phrases of length $D = 3$, each indicated by a pink square in the musical staff. We obtain the distribution of different not duration patterns for each of the phrases, where in this case only the patterns of the first two phrases repeat. The heterogeneity in this case is $\mathcal{H} = 0.405$.

## Visibility graph

In this method, we use the network generation algorithm developed by Lacasa et al. in [3]. The fundamental idea of this algorithm is to map a time series into a graph by following a simple rule: considering the height of each the points of a time series represented as a column, link every column with those that can be seen from the top of the considered one, obtaining an associated graph (see Figure 3a). On a perceptual level, the length of sound events plays a decisive role when it comes to understanding a sound structure: "[...] Lengthening a sound does not have the same effect when the sound is small and when it is large. If it is small, the longer sound plays the role of an accentuated element and determines the beginning of the grouping; if it is large, the longer sound plays the role of an interval between two patterns and terminates the grouping [...]" [7]. Here we are talking about the length of a sound, but it is intuitively clear that a set of sounds followed by a long silence is also perceived as a group. Thus, the visibility graph obtained from the time series of the rhythmic length values of the sound events in a given musical piece, clusters together short-time sound events and separates such clusters using the long-time ones (see Figure 3a), which holds a good correspondence with the perceptual behavior presented before.

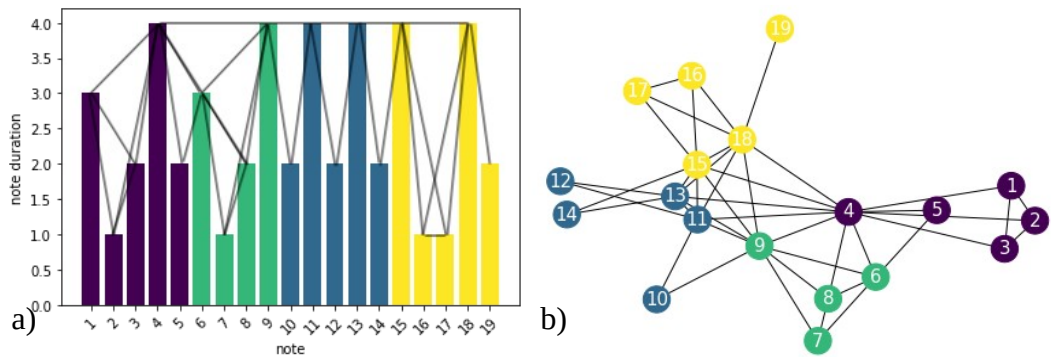

Figure 3. a) Time series of the extract in Figure 1 and b) the visibility graph generated with the communities obtained by Louvain's algorithm.

To find the *parts* in which the piece can be divided, we apply a community detection algorithm to the visibility graph. Here, we use Louvain's clustering algorithm, which groups a set of nodes into a community if the density of edges inside the community is higher than what we would get by chance if we randomly assign edges according to the degree of the nodes [6]. In Figure 3, we observe that some nodes of the graph play an ambiguous role at the beginning and the end of a cluster; this makes the community algorithm fail to put together all the notes in measure 2 (see node number 10, compared to number 5). Despite the ambiguities in the final or initial nodes in the communities, the graph and time series in Figure 3 show an accurate correspondence with the structure of the initial phrase in Mozart's sonata; the first two communities are the thematic motiv, one in the tonic and one in the $V_6$ dominant function (in purple and green), the third community is the transition to the V dominant function (in blue), and the last community is the cadence to the V dominant function (in yellow).

The graph topology contains the clustering information of the piece, and we can extract such information using graph metrics. For instance, we focus on the *degree correlation,* which characterizes a graph by comparing the number of links $k_i$ of a given node *i* (such number is called the *degree of i*) with the average degree $k_{nn}$ of the nodes that *i* is connected to. Summing over all *i* and normalizing, we end with a function $k_{nn}(k)$, called the *degree correlation* of the graph, which is the average degree of the neighbors of all degree-*k* nodes [11]. This function quantifies how the low-degree and the high-degree nodes are linked among each other. A standard way to analyze this property is by approximating $k_{nn}$ with a power-law distribution so that $k_{nn}(k)=ak^b$. Note that if we use a *log-log* plot, the power-law distribution takes a linear form with a given slope $b^1$. The sign and magnitude of this slope characterize the graph behaviour. A more general global metric that characterizes the degree correlation of a graph is known as the *assortativity coefficient* or *Pearson correlation coefficient* which quantifies the difference between the degree correlation of the graph and a graph with the same amount of edges and vertices but distributed following a normal degree distribution [11]. We denote the assortativity coefficient by the letter *r,* and it takes real values between -1 and 1. The values of *r* close to 1 indicate an *assortative* behaviour, meaning that high-degree nodes tend to be connected to other high-degree nodes, while low-degree nodes are mostly linked with low-degree nodes. In contrast, when *r* is close to -1, we have *disortative* behaviour, which is the opposite. If the value of *r* is close to 0 then the graph assortativity behaviour is said to be *neutral*. In fact, small values of *r* indicate that (low) high-degree nodes do not have a clear preference to be linked with other (low) high-degree ones but with nodes of arbitrary degree instead. In terms of the visibility graph of the rhythmic content of a piece, these metrics allow us to measure how long-duration sound events are correlated with short-duration ones, in the sense that the distribution of events permits or not to link consecutive notes.

---

1 Depending on the degree distribution of the graph, other definitions for this coefficient can be given. For more information see [11].

In this sense, the graph *transitivity* is another metric that indicates whether the distribution of sound events permits to link consecutive notes or not. Mathematically, the transitivity quantifies the proportion of *triangles* in a graph with respect to *all the possible triangles* with the same amount of nodes. A triangle exists if, given three nodes in a graph, there is an edge connecting each pair of nodes. For example, a fully connected graph has a transitivity equal to 1 because every triad of nodes forms a triangle while a tree-like graph has a transivity equal to 0 because it has no loops and hence no triangles. In the visibility graph, short-duration events tend to form triangles with long-duration ones, and in some cases, between them, depending on the rhythmic time-series distribution. In the *Results* section, we will show an example of these analyses.

## Dynamic and static metrics

We can take two distinct approaches when computing any of the complexity metrics we describe above: a *static*, global indicator of the musical piece's rhythmic complexity or a *dynamic*, local measure of the parts of the piece. The static approach is easier; we only need to compute the IOI vector, apply the formulas and obtain a single number for the piece for each of the complexity metrics. This way, we can compare the complexity of two scores as long as we use the same parameters for both. Score A is quantitatively more *heterogeneous* than score B if the *permutation entropy* of A is higher than that of B. The dynamic approach consists of taking a time window $W$ of a fixed size, say, $W = 4$ measures, and computing the complexity metrics inside $W$. Then, we slide the window forward for some time interval $\delta W$, say, $\delta W = 1$ measure, and compute the metrics again. We continue this process until the end of the score is reached. This results in a time series where its *i*-th element is the complexity of the piece's between measures $i\delta W$ and $i\delta W + W$.

The global indicator of a complexity score is *not* the same as the average of its dynamic counterpart. This complements our idea of form as an emergent property of rhythm, and while it is true mathematically, it also makes sense musicologically. Imagine a score with complex musical motifs, but these motifs repeat a lot during the piece without any significant alterations. Thus, the dynamic analysis of the music would reveal a high rhythmic complexity. In contrast, the global complexity would be low given that the motifs, although complex in their own right, become predictable and repetitive in the long run.

## RESULTS

In this section, we analyze the complete first movement of Mozart's Sonata using our complexity metrics. As a form of comparison, we perform a traditional analysis of the score and show it in Figure 4. In short, the score consists of an AABB structure, where A presents the theme with a dominant half-cadence and then repeats it, resolving to the tonic. Section B also consists of two subparts; the first one is a variation of the first measure of the theme that plays the role of the development, finishing with a half-cadence to the dominant, and the second one is the re-exposition (or recapitulation) of the main theme with a prolongation of the final cadence to the tonic.

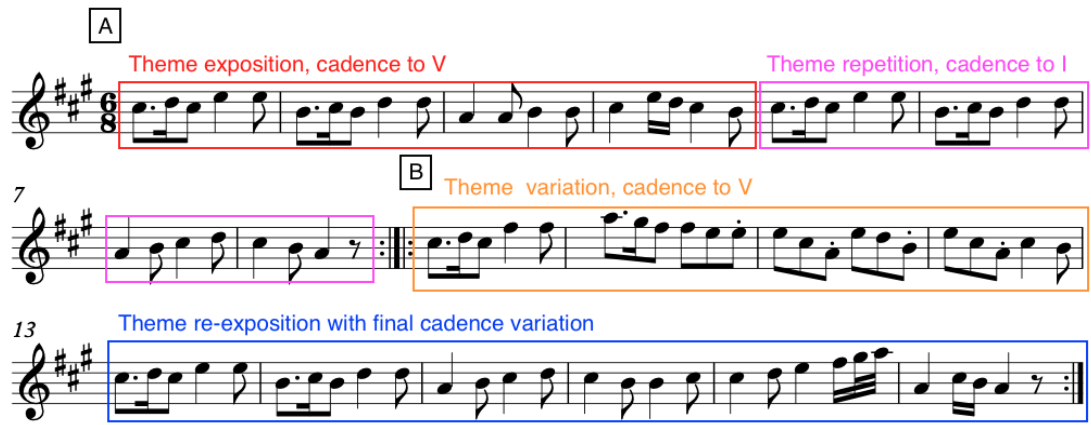
Figure 4. Formal analysis of the first movement of Mozart's Sonata.

In Figure 5, we show the results of both the dynamic and static metrics of heterogeneity and syncopation for Mozart's Sonata. For the dynamic analysis, we take a time window size of $W = 2$ measures[2] so that *i-th* element of the series indicates the complexity between measures $i$ and $i + W$. For each section, we find time-series motifs that repeat exactly whenever a section repeats. Moreover, while the motifs are not exactly the same for different sections, they share several features. These motif features consist of starting with high complexity (due to the ♪ ♪♪ ♩ ♪ or equivalent melodic cells), then getting lower (due to the ♩ ♪♩ ♪ or equivalent musical cells), and rising again due to a variation of the first musical cell. This motivic structure is present throughout all the sections of the piece. However, according to our metrics, section A1 is more complex than A2, Section B1 has a similar complexity to section B2, while, compared with section A, the "getting lower" cell is much less complex than in section A. This kind of analysis lets us understand the local structure of the piece, compare the different sections in terms of their complexity, and recognize rhythmic patterns, musical cells, and time series motifs at different scales.

Regarding the static complexity indicators for Mozart's Sonata, we observe that the global syncopation (0.11) is very similar to the average syncopation of the dynamic analysis (~0.1). Nevertheless, the global heterogeneity (0.65) is significantly higher than the average heterogeneity of the time series (~0.4), indicating that, overall, the variation of the musical cells is frequent and noticeable enough for the permutation entropy to be higher than the sum of its parts in the overall picture of the piece.

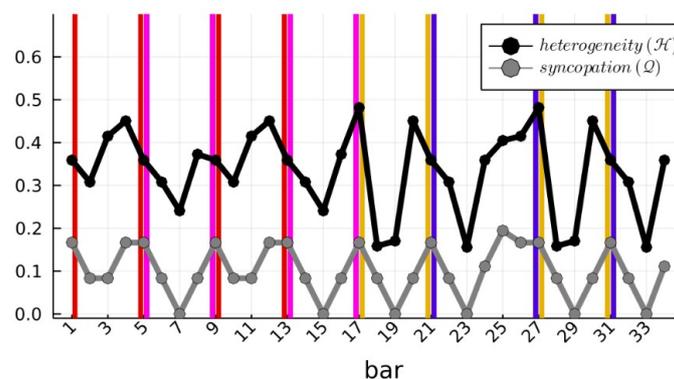
Figure 5. Dynamic metrics for the *rhythmic heterogeneity* (black) and the *syncopation* (gray) of Mozart's Sonata. We color the solid vertical lines according to the formal analysis in Figure 4.

---

2 The window size $W$ and the sliding window $\delta W$ are design parameters of the dynamic analysis, and we will make a systematic analysis of their effect in future work.

In terms of the visibility graph, we obtain communities that accurately match the traditional formal analysis of the piece, as we show in Figure 6. Note that the formal subsections of the first four measures in the last section are no longer present in the communities. Instead, all section A is clustered together. This happens because a larger node at the end of this section corresponds to the quarter note plus the octave silence that follows it in measure 8. In section B, the communities correspond with the subsections given by the traditional analysis exactly.

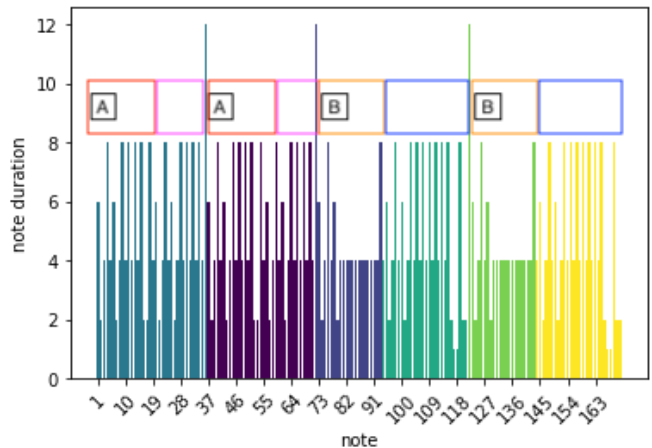

Figure 6. Time series of the rhythmic values and the communities found. The formal analysis of Figure 4 is also shown inside the plot.

We found that the *degree correlation* of the visibility graph of Mozart's Sonata has assortative behaviour, given that the power-law approximation has a positive sign and the assortative coefficient is positive (see Figures 7 and 8). This result shows that events with a long duration (in this case the quarter notes) will *see* almost every other note in the corresponding visibility graph, so that quarter notes will have a high degree and will be connected to other events with a high degree. On the other hand, short-duration events are *trapped* between long-duration ones, so their resulting degree is low in general.

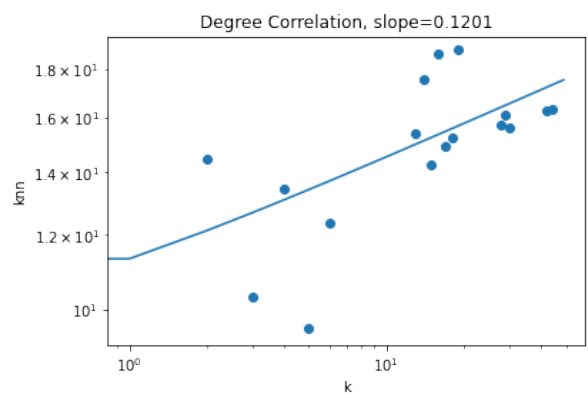

Figure 7. Degree correlation of the visibility graph with a power-law approximation coefficient of $b = 0.1201$.

**Static analysis of complexity for scores of different musical periods**

As the last example of what we can do with our metrics, we show in Figure 8 a static analysis of several musical pieces from the Baroque, with Bach's Gavotte en Rondeaou BWV1006, Classic, with Mozart's mentioned Sonata 11, and Haydn's Sonata in F Hob. XVI-23, the Romantic, with Chopin's Nocturne op. 9 no. 1, and the Impressionist, with Debussy's Prelude no. 8 from book 1.

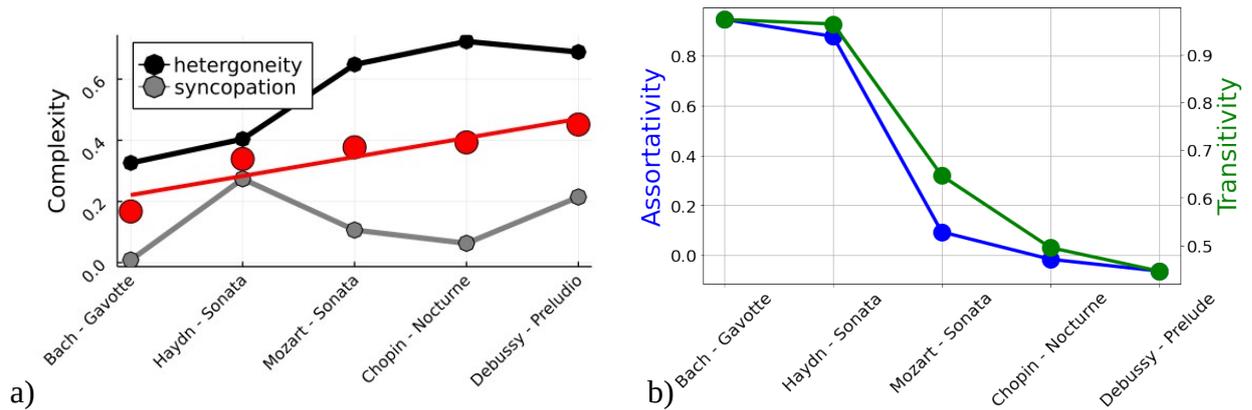

Figure 8. Global metrics of rhythmic complexity for different, time-ordered musical periods: Baroque, Classical, Romantic, and Impressionist. a) Rhythmic heterogeneity (black) and syncopation (gray); b) assortativity coefficient (blue) and transitivity (green) of the visibility graphs.

In Figure 8a, we order from lowest to highest average rhythmic complexity and find that such an order coincides with the period to which they belong. Except for the Impressionist period, the heterogeneity (permutation entropy) gets higher for each period. The syncopation oscillates over time, but we find that, in general, a decrease in heterogeneity comes with an increase in syncopation. In Figure 8b, we order from highest to lowest the assortativity and the transitivity of the piece's visibility graph, showing an inverse behaviour with respect to the complexity. These results show that the highest assortativity and transitivity may be correlated with high rhythm regularity. However, the number of pieces analyzed here is still very small to derive robust conclusions about the rhythmic complexity of different musical periods. Nevertheless, when one looks into these pieces in detail, the complexity values obtained make sense, which is a promising result.

## CONCLUSION

In this paper, we propose three metrics of the rhythmic complexity of a musical piece: one that captures the heterogeneity of the rhythms, one that captures their syncopation, and one that captures its subjective groupings throughout the piece. These metrics show a significant correspondence with the traditional music form analysis, as we showcase the first movement of Mozart's Sonata. As these methods *find* the form with simple quantitative rules of classification, not with a preconceived notion, they help to generate a new way of understanding this concept as an emergent property of rhythm. Forthcoming work will show how our methods work on larger compound pieces, as well as non-Western and post-1900s Western music.

We construct these metrics using the *inter-onset durations* between each note in the piece, and we propose two ways of using these metrics: a global, static approach in which the whole musical piece has an associated complexity number, and a dynamic, local approach in which we generate a time series of the piece's complexity. We can obtain the average local complexity by taking the average of all the time series values. Interestingly, the average local complexity differs from the static global complexity, which suggests that form emerges from the various local interactions in a piece. Both of these approaches offer innovative insightful ways to study rhythm in music.

Finally, we stress that 1) the inter-onset durations vector is not the only way to understand rhythm, and 2) rhythm is not the way to understand form. About the first point, we suggest that one could incorporate *silence* into the notion of rhythm in future work. If we understand silence as a sound event on its own, several notions of complexity and form will have a different interpretation. About the second point, we believe that to understand the actual complexity of a musical piece, we need to incorporate other elements that constitute a music piece: melody, harmony, timbre, and interpretation. This is without mentioning the weight of culture, historical period, and socioeconomic circumstances surrounding the artists and their art.